\documentstyle[11pt,aaspp4,psfig]{article}

\slugcomment{}

\lefthead{}
\righthead{}

\begin{document}

\title{Reconstructing Galaxy Spectral Energy Distributions from
Broadband Photometry}

\author{I. Csabai\altaffilmark{1}, A. J. Connolly\altaffilmark{2},
A. S. Szalay\altaffilmark{1}}

\affil{Department of Physics and Astronomy, The Johns Hopkins
University, Baltimore, MD 21218\\ Electronic mail:
csabai@skysrv.pha.jhu.edu, ajc@skysrv.pha.jhu.edu,
szalay@skysrv.pha.jhu.edu}

\and

\author{T. Budav\'{a}ri\altaffilmark{3}}
\affil{Department of Physics, E\"{o}tv\"{o}s University,
Budapest, Pf. 32, Hungary, H-1518}

\altaffiltext{1}{Department of Physics, E\"{o}tv\"{o}s University,
Budapest, Pf. 32, Hungary, H-1518}
\altaffiltext{2}{Department of Physics and Astronomy, University of Pittsburgh, Pittsburgh, PA 15260}
\altaffiltext{3}{Department of Physics and Astronomy, The Johns Hopkins
University, Baltimore, MD 21218}

\begin{abstract}

We present a novel approach to photometric redshifts, one that merges
the advantages of both the template fitting and empirical fitting
algorithms, without any of their disadvantages. This technique derives
a set of templates, describing the spectral energy distributions of
galaxies, from a catalog with both multicolor photometry and
spectroscopic redshifts. The algorithm is essentially using the shapes
of the templates as the fitting parameters. From simulated multicolor
data we show that for a small training set of galaxies we can
reconstruct robustly the underlying spectral energy distributions even
in the presence of substantial errors in the photometric observations.
We apply these techniques to the multicolor and spectroscopic
observations of the Hubble Deep Field building a set of template
spectra that reproduced the observed galaxy colors to better than
10\%. Finally we demonstrate that these improved spectral energy
distributions lead to a photometric-redshift relation for the Hubble
Deep Field that is more accurate than standard template-based
approaches.

\end{abstract}




%

\section{Introduction}

With the application of photometric-redshifts to wide-angle,
multicolor photometric surveys the study of galaxy evolution has moved
from expressing the evolution as a function of observable parameters
(e.g.\ magnitudes and colors) to one where we can describe the
evolution of galaxies in terms of their physical attributes (i.e.,
their redshift, luminosity and spectral type). Over the last several
years with new multicolor surveys coming on-line these techniques have
become increasingly popular enabling large, well defined statistical
approaches to galaxy evolution.

In the astronomical literature there are a number of different
approaches to estimating the redshifts of galaxies from their
broadband photometry.  While these techniques differ in their
algorithmic details they share the same underlying goal. We wish to
model the change in galaxy color as a function of redshift (and galaxy
type) and use these models to estimate galaxy redshifts, in a
statistical sense. For simplicity we divide the differing techniques
into two classes, those which use spectral energy distributions
(whether derived from models or empirically from observations of local
galaxies) as spectral templates and those which derive a direct
empirical correlation between color and redshift using a training set
of galaxies.

Template based photometric-redshifts are constructed by comparing the
observed colors of galaxies to a set of galaxy spectral energy
distributions. Their strength is that they are simple to implement and
can be applied over a wide range in redshift. Their main limitation is
that we must know the underlying spectral energy distributions of
galaxies within our sample. Comparisons of the colors of galaxies with
spectral synthesis models (Bruzual and Charlot 1993) have shown that
the modeling of the ultraviolet part of galaxy spectra is highly
uncertain (whether this is due to uncertainties in the modeling of the
stars or due to the effect of dust is unclear). Consequently,
photometric-redshift estimates are most accurate when we apply
empirical spectral energy distributions derived from observations of
local galaxies (e.g.\ Sawicki et al 1997). These empirical relations
are however constructed from only a handful of galaxies that have been
observed in detail and there is no guarantee that they represent the
full distribution of galaxy types (particularly when we include the
effects of evolution with redshift).

The second approach is to derive a direct correlation between the
observed colors of galaxies and their redshifts using a training set
that contains spectroscopic and photometric data. The strength of this
technique is that the relation is purely empirical; the data
themselves define the correlation between color and redshift. The
effects of dust and galaxy evolution that are present in the training
set are, therefore, implicit within the derived correlation. Their
weakness is that the correlations cannot be extrapolated to redshifts
beyond the limits of the training set and that a sample of galaxies
with redshifts (and selected to have a broad color distribution) must
be present before the photometric-redshift relation can be derived.

Clearly if we can combine these two approaches we may derive the
optimal approach for estimating galaxy redshifts. If we can use a
training set of galaxies to define the underlying spectral energy
distributions (which will include the effects of evolution and dust)
then we can apply these empirical template spectra over a wide range
in redshift. In this paper we describe a fundamentally new approach to
photometric-redshifts that extends our previous work on estimating
galaxy redshifts from broadband colors such that we construct spectral
energy distributions directly from the broadband data. In Section 2 we
outline the physical and mathematical basis of this new approach. In
Section 3 we apply these techniques to a sample of galaxies with
simulated colors, showing that we can recover the underlying spectral
energy distributions that describe these galaxies.  Section 4 applies
the optimization procedure to the multicolor photometric data of the
Hubble Deep Field (Williams et al 1996) and shows that this technique
can be used to improve the accuracy of template-based photometric
redshift relations.  Finally in Section 6 we describe the application
of these techniques to analysis and the modeling of galaxy spectral
energy distributions.

\section{Building Spectral Templates from Broadband Photometry}

In an earlier work (Connolly et al 1995a) we described how to model the
empirical correlation between the colors of galaxies and their
redshifts by fitting a multi-dimensional polynomial relation. This
technique proved successful for estimating the redshifts of galaxies
in the $0<z<1$ regime but was built on a very general but somewhat
unphysical basis that the color-redshift relation can be described by
a low order polynomial. Ideally we want the underlying basis on which
we define the photometric-redshift relation to be physically
motivated. If we can construct a set of low resolution spectral energy
distributions directly from a set of galaxies with multicolor
photometry then we can achieve this goal (an empirical, physical
basis).

If we consider a galaxy, at a redshift $z$, observed through a series
of broadband filters then the restframe fluxes observed through each
filter, $f_{k}$ can be described by,
\begin{equation} 
f_{k} = \int R_{k}(\lambda ) S(\lambda / [1+z])d\lambda
\end{equation}
where $R_k(\lambda )$ is the response function of the $k$th filter and
$S(\lambda / [1+z])$ is the spectral energy distribution of a galaxy
blueshifted to the galaxy's restframe. The response function,
$R_k(\lambda )$, includes not only the filter transmission but also
the instrumental and observational effects such as the CCD quantum
efficiency and the change in effective shape of the filter due to
absorption by the atmosphere. The spectral energy distribution,
$S(\lambda / [1+z])$, is the true underlying spectrum that includes
the stellar composition of the galaxy together with the effect of
intragalactic extinction. For high redshift objects the effects of 
the IGM should be built in the above equation. 

We can see that $f_{k}$ is nothing more than a convolution of the
input spectrum with the filter response function. Thus from a
photometric catalog of galaxies with identical restframe spectral
energy distributions, given the filter response functions, we can
deconvolve the underlying spectral energy distribution. More exactly
we can recover a low resolution slice of the spectrum whose limits are
defined by the wavelength range over which the filters extend and the 
redshifts of the galaxies in the catalog.

If we observe just one object there are, of course, a wide range of
spectra that could give the exact flux values passing through the
broadband filters. The question then arises, if instead we have an
ensemble of $N$ galaxies over a range of redshift, with accurate
redshifts and $K$ multicolor photometric observations per galaxy, can we invert
this relation to recover the underlying spectral energy distributions
(even in a statistical sense)? In principle, for a sample of galaxies
spread over a range in redshift, we have $N\times K$ measurements of
the underlying spectral energy distributions. For a series of optical
and near-infrared filters the resolution of our reconstructed spectra
would be proportional to the rest wavelength range sampled by the
filters divided by $N\times K$.

The advantage of this technique is that it is numerically
straightforward to calculate. The deconvolution algorithms do not
require a large amount of computational power. Its main weakness is
that in the real world we cannot construct a large catalog of galaxies
with identical spectral energy distributions on which to apply the
deconvolution algorithm. We can circumvent this, however, by noting
that galaxy spectra can be described by a small number of
orthogonal spectral components or eigenspectra (Connolly et al
1995b). Each galaxy spectrum can be written as,
\begin{equation} 
S(\lambda)=\sum _{j=1}^{J} a_{j} E_{j}(\lambda )
\end{equation}
where $E_{j}(\lambda)$ are the $J$ eigenspectra and $a_{j}$ are the
expansion coefficients. Each galaxy spectral type can then be
described by a linear combination of the eigenspectra (i.e.\ only the
expansion coefficients, $a_{j}$, differ as a function of spectral
type).  From observations of local star forming and quiescent galaxies
it has been shown that the number of components (or eigenspectra)
required to reconstruct the continuum shape of a galaxy spectrum (to
an accuracy of better than 1\%) is small, typically 2--4 (Connolly et
al 1995b).

Utilizing the eigenspectra formalism simplifies the deconvolution
problem in two ways. By restricting the number of components we need
to reconstruct from the multicolor data to 2--4 the inversion process
is much simplified. Secondly, the fact that any galaxy spectrum can be
described by this small number of components enables us to use all of
the available data (i.e.\ we do not have to restrict our analysis to a
single class of galaxies with similar spectral types).

\subsection{Template spectrum estimation}

The goal of the deconvolution is to provide a set of $J$ eigenspectra,
$E_j(\lambda)$, that best represent the observed colors of the galaxy
training set. We can parameterize the eigenspectra as a linear
combination of basis functions such that,
\begin{equation} 
E_{j}(\lambda )=\sum_{l=1}^L b_{jl} B_{l}(\lambda),
\end{equation}
where $B_{l}(\lambda)$ are a set of $L$ basis function vectors (e.g.\
Legendre polynomials) and $b_{jl}$ are their relative expansion
coefficients. In this paper we choose to parameterize the eigenspectra
in terms of a linear combination of Legendre polynomials. The choice
of the parameterization is, however, completely arbitrary. There are
many ways we could describe the galaxy eigenspectra.  In the simplest
case the coefficients $b_{jl}$ could be just the flux values of the
eigenspectra measured at a fixed set of wavelengths (i.e.\ the the
basis functions would be delta functions centered at these
wavelengths). We would then be reconstruct the eigenspectra directly
rather than expressing it in terms of a linear combination of
functions (Budavari et al 1999).  We also note that, while we use the
term eigenspectra, the above procedure does not guarantee (nor is it
necessary) that the eigenspectra be orthogonal.

From Equations 1, 2 and 3 we can now estimate that color of a galaxy
in terms of the linear combination of eigenspectra or basis
functions. For the $i$th galaxy in a ensemble of photometric
observations the estimated flux through the $k$th filter is given by,
\begin{eqnarray} 
f_{ik}^{e} &=& \int _{0}^{\infty }R_{k}(\lambda ) \sum_{j=1}^J a_{ij}
E_{j} (\lambda / [1+z_{i}]) d\lambda\\
           &=& \int _{0}^{\infty }R_{k}(\lambda ) \sum_{j=1}^J \sum_{l=1}^L 
	a_{ij} b_{jl} B_{l}(\lambda / [1+z_{i}]) d\lambda
\end{eqnarray}

We can now define a $\chi^2$ or cost function that describes the
distance between the observed flux values, $f_{ik}^{m}$, measured for
particular galaxy and those predicted by the eigenspectra,
$f_{ik}^{e}$. We write the form of the cost function as a $\chi^{2}$,
weighted by the measured flux errors $\sigma _{ik}$ but other
distances can be also used, dependent on how one would like to weight
the different observations.
\begin{equation} 
\chi^2 = \sum _{i=1}^N \sum _{k=1}^K \frac{[f_{ik}^{e}-f_{ik}^{m}]^{2}}{\sigma
^{2}_{ik}}. 
\end{equation}
The cost function depends on parameters $ a_{ij} $ and $ b_{jl} $. The
minimum of this cost function determines the set of optimal parameters
(in other words eigenspectra and expansion coefficients) that give the
best estimation of fluxes in this framework. Of course, the larger the
catalog of galaxies with multicolor observations the more
non-redundant parameters we can optimize for and, consequently, the
finer the resolution in the eigenspectra.

By carefully choosing how we generate the eigenspectra we can have a
cost function whose minimum can be found almost analytically. The
variable parameters of the cost function will be $a_{ij}$ and
$b_{jl}$.  Since Equation 5 is linear in both of them they will appear
in a quadratic form in the $ \chi^2 $ cost function. At the $\chi^2$
minimum all of the derivatives of the cost function should be zero. If
we consider the values $a_{ij} $ as constants, the equations with the
derivatives in $b_{jl} $ will give a set of $J\times L$ linear
equations.  In similar way keeping $b_{jl}$ constant we will have
$N\times J$ linear equations for $a_{ij}$, or to be more exact it
breaks up into $J$ sets of $N$ linear equations with $N$ unknowns in
each. Each of these set of linear equations can be solved
independently. Therefore, by iteratively solving the two sets of
linear equations (holding one set of coefficients constant while
solving for the other coefficients) one can minimize the cost function
in a efficient manner.

In fact one could generate the eigenspectra in many other ways, for
example, using a $tanh()$ function to mimic the 4000 \AA\ break and a
power law function to represent the star formation at the ultraviolet
end of the spectrum. Parameters could occur in the cost function in
more complex forms than showed above. If the variable parameters are
not in a quadratic form in the cost function nonlinear optimization
methods (such as different types of gradient descent methods or
simulated annealing if local minima cause problems) could be used
instead of solving linear equations. This would be computationally
harder but the shape of the eigenfunctions would not be restricted to
a particular basis.

The limitation on the number of parameters and, therefore, the number
of Legendre polynomials, is defined by the set of galaxies with
multicolor photometric observations. As described previously, if we
have $N$ galaxies with observations in $K$ passbands then we have $N
\times K$ independent measurements. If we want to describe the
distribution of galaxy types by $J$ eigenspectra (in our case 3
eigenspectra) then we have $J \times N$ constraints. This means that
the number of degrees of freedom in the system and, therefore, the
maximum number of parameters (or polynomials) we can solve for, $X$,
is,
\begin{equation}
X \le \frac{N(K-J)}{J}
\end{equation}

As we noted earlier the parameterization we choose to describe the
eigenspectra by is completely arbitrary. A second, and more physical,
way to visualize how the number of observations (number of galaxies
and passbands ) relates to how well we can recover the underlying
eigenspectra is to think of the parameters $X$ as the number of
wavelengths at which we can sample a galaxy spectrum. This involves
using the values $E_{j}(\lambda _{l})$ of a discretely sampled low
resolution eigenspectrum as optimization parameters. The basis
functions will, therefore, have a constant value 0 except at a
selected $\lambda_{l}$ where the value is 1. In such a way we have a
direct, low resolution realization of the underlying eigenspectra that
describe the observed galaxy population. It is then clear that, given
the wavelength interval we wish to reconstruct, $X$ defines the maximum
resolution of the resultant eigenspectra.

The number of observations and the passbands are given by the measurements 
but we have some freedom to chose the number of eigenspectra and the resolution.
For example if we want to use the templates for photometric redshift estimation 
we have to have enough resolution to reconstruct the broadband features of 
the spectrum. So we need resolution at least the width of the (blueshifted) 
filters. On the other hand to represent all different spectral types even 
rare ones we would like to have large number of templates. If the number of
observations is not enough to satisfy both of these requirements one has to
chose between representing equally the spectra of all of the objects but with  
poor quality or to represent the typical ones with better resolution and have 
a few outliers. Comparing the final value of the cost function or the 
quality of photometric redshift estimation can help to set these parameters
optimally.

\section{Application to Simulated Data}

To demonstrate the validity of this technique and to determine the
accuracy to which we can reconstruct galaxy spectral energy
distributions from broadband photometry we initially apply the
algorithm to a set of simulated data. From the Bruzual and Charlot
spectral synthesis models (BC96, Bruzual and Charlot 1995) we
construct a set of galaxy spectra using a simple stellar population
ranging in age from 0 yrs to 20 Gyrs (with solar metalicity and a
single burst of star formation). In total the sample contains 222
spectra covering the spectral range 200 \AA\ to 2.2 $\micron$.  From
these spectra we apply a Principal Component Analysis or
Karhunen-Lo\`{e}ve transform (Karhunen 1947, Lo\`{e}ve 1948, Connolly
et al 1995b) to construct a series of orthogonal eigenspectra. The
first two of these eigenspectra are shown in Figure 1a and Figure 1b
respectively.

Using the first two eigenspectra we simulate the colors of galaxies as
a function of redshift. The expansion coefficients $a_i$, or mixing
angles, are designed to produce a set of galaxy colors that match the
color distribution of galaxies within the local Universe. In total, we
construct a sample of 616 galaxies with U, B, V, I, J, H and K
photometry covering the redshift range $0<z<1.35$.  The upper limit on
the redshift range is imposed for two reasons, to match the redshift
distribution of those galaxies in the Hubble Deep Field with $V_{606}
< 24$, and secondly to avoid the added complication of including the
attenuation due to the intergalactic medium (e.g.\ Madau et al 1996).

\subsubsection{Reconstructing Bruzual and Charlot Spectra}

Given these simulated data (colors and redshift) we use the algorithms
described in Section 2 to reconstruct the underlying eigenspectra.
The redshift distribution of the input catalog of galaxies defines the
wavelength range over which the eigenspectra can be reconstructed (the
upper and lower bounds being defined by the restframe upper and lower
filter cutoffs of the K and U passbands respectively). For the
redshift range of the Bruzual and Charlot data we can recover the
spectral energy distribution over a wavelength interval of
approximately $950<\lambda<22000$ \AA.

Initially we sample the eigenspectra that we wish to recover in 20
bins (sampled linearly in wavelength), with a spectral resolution of
approximately 1000 \AA. Each of the two approaches (i.e.\ using
Legendre polynomials and evenly sampled random spectra) outlined in
Section 2 were applied to the data and were found to give identical
results. In the section below we will, therefore, only discuss the
optimization of the eigenspectra using Legendre polynomials. A
subsequent paper (Budavari et al. 1999) will discuss in more detail
techniques that can be applied directly to the spectra themselves.

For basis functions, $B_{l}(\lambda )$, we used the first $ 20 $
Legendre polynomials.  The optimization procedure was started with the
Legendre polynomials, $B_{l}(\lambda )$, specified by a series of
random numbers. The optimization was found to converge rapidly for the
616 galaxies within our simulated data. After 50 iterations, a few
minutes of workstation cpu time, the cost function was found to be
stable (varying by less than 0.1\% from one iteration to the next).
The rapid convergence of the relative error as a function of iteration
is shown in Figure 2.

Comparison between the eigenspectra input into the simulations and the
spectra reconstructed by our optimization technique is not
straightforward. While the spectral templates that the reconstruction
technique derives should occupy the same subspace as the original
eigenspectra there are many non-unique ways to achieve this (i.e.\ the
output spectra can be a rotated subset of the input spectra).  To
transform the input and output spectral templates into a common form,
for each input eigentemplate, we calculate the linear combination of
the output eigentemplates that gives the closest representation of
input eigenspectrum (i.e.\ we project the original eigenspectra
onto the reconstructed eigenbasis). This produces a set of eigenspectra
that can then be directly compared with the BC96 input basis. In
Figure 1 we show a comparison between the BC96 eigenspectra and those
we derive from the optimization technique. Clearly there is an almost
perfect one-to-one match between the two with the rms scatter less
than 7\%  both for the first and second eigenspectra.

The analysis we show in Figure 1 is the ideal case (where the noise in
the observations is negligible). As deep photometric and spectroscopic
surveys tend to push analyses of the data to the limit of the survey
we need to investigate the effect of photometric uncertainties on the
reconstruction of the underlying spectral energy distributions. For
simplicity we assume a constant photometric error across all passbands
(i.e.\ we do not allow for the lower signal-to-noise that are
prevalent in ultraviolet observations of intermediate redshift
galaxies).

Figure 3 shows the effect of increasing the photometric uncertainty on
the reconstruction of the first eigenspectrum. The solid line is the
reconstructed eigenspectrum with no noise added, the triangles,
squares, and circles show the effect of adding 5\%, 10\% and 20\% flux
errors respectively. As we see even with very low signal-to-noise data
the eigenspectra can be reproduced to a very high accuracy. The large
number of galaxies present within the sample means that each spectral
interval (i.e.\ the 1000 \AA \ spectral bins) is sampled by multiple
galaxies. The coaddition of these multiple realizations increases the
signal-to-noise of the reconstructed spectrum (relative to the input
data).

We note that for a flux error in excess of 5\% the long wavelength
end of the eigenspectrum becomes significantly more noisy than the
remaining spectral regions. This arises because the longest rest
wavelengths are only sampled by the lowest redshift galaxies.
Therefore, the reconstructed spectrum will have a larger uncertainty
where the spectral values are constrained by only a small number of
data points. For decreasing signal-to-noise the longest wavelength
spectral regions will be the most susceptible to the effect of the
noise and can, therefore, be used as an indicator of when photometric
uncertainties become significant within an analysis.

\subsubsection{Photometric redshifts from empirical eigentemplates}

Having reconstructed the eigenspectra that describe the distribution
of galaxy colors we utilize these spectra to derive a photometric
redshift relation. We note that the reconstruction technique does not
involve minimizing the difference between the spectroscopic redshift
of a galaxy and its photometric redshift rather it minimizes the
differences in the observed and estimated colors. This means that the
dispersion about the photometric redshift relation is an accurate
measure of how well the reconstructed spectra match the simulated
data.

We apply the standard template-based photometric redshift relation,
adapted to utilize eigenspectra (e.g.\ Benitez 1999). For the range of
redshifts we wish to consider (in the case of our simulations
$0<z<1.35$) we define a redshift dependent $\chi^2(z)$,
\begin{equation}
\chi^2(z) = \sum_{i=1}^N \sum_{k=1}^K \frac{(f_{ik}^m - \sum_j a_j
E_{jk}(z))^2}{\sigma_{ik}^2}
\end{equation}
where $f_{ik}^m$ is the color of galaxy $i$ observed through the $k$th
filter, $\sigma_{ik}$ is the flux error, $a_j$ are the expansion
coefficients of the eigensystem (which we also solve for) and $E_{jk}$
is the color of the $j$th eigenspectrum observed through the $k$th
filter. Minimizing this relation gives the estimated redshift of the
galaxy (and is a simple and fast 1 dimensional problem).

While the deconvolved eigentemplates are found to be marginally
susceptible to the effects of photometric uncertainties within the
data we find that the dispersion in the resulting photometric redshift
relation derived from these spectral templates is sensitive to the
photometric noise. In Figure 4 we show the dispersion in the
photometric redshift relation for a set of simulated data with
photometric uncertainties of 0\%, 5\%, 10\% and 20\%. For each sample
the eigenspectra were derived directly from the data themselves (i.e.\
including the photometric errors). These eigenspectra were then used
to derive the photometric-redshift relation. The top left panel shows
the photometric-redshift relation in the presence of no noise
($\sigma_z = 0.018$), the top right the effect of 5\% flux error
($\sigma_z = 0.049$), the bottom left the effect of 10\% flux errors
($\sigma_z = 0.104$), and the bottom right the effect of 20\% flux
errors ($\sigma_z = 0.258$).

The correlation between the dispersion about the photometric-redshift
relation and the signal-to-noise of the data is not, however, due to
errors present within the reconstructed eigenspectra. We demonstrate
this in Figure 5. We derive the eigenspectra for a sample of galaxies
with 20\% flux errors (i.e.\ relatively low signal-to-noise). We then
use these eigenspectra in a photometric-redshift analysis of a set of
data with no flux errors. The resultant relation has a dispersion of
0.051 (substantially smaller than the 0.25 derived from the
simulations with 20\% flux errors). The limitation on the accuracy of
a photometric redshift is, therefore, almost entirely due to the
signal-to-noise of the data set to which we wish to apply the
relation.

\section{Deriving Spectral Energy Distributions from Broadband Photometry}

\subsubsection{Templates from Multicolor Photometric Observations}

The most natural data set on which to apply our deconvolution
techniques are the Hubble Deep Field observations (HDF; Williams et al
1996). These data comprise a single WFPC2 field observed in four
ultraviolet/optical passbands (F300W, F450W, F606W and F814W). The HDF
has been the target of a substantive effort to obtain deep
spectroscopic redshifts and follow up near-infrared and longer
wavelength imaging for the central and flanking fields. It represents
one of the densest regions of the sky in terms of published multicolor
photometry and spectroscopy. In total there are over 110 galaxies
within the HDF with high signal-to-noise optical and near-infrared
colors and redshifts.

\subsection{Template estimation}

To enable a direct comparison between the accuracy of
photometric-redshifts based on our template estimation algorithm and
those derived by others from standard techniques we use the
photometric catalog of Fernandez-Soto et al (1999). These data are
based on the optical WFPC2 colors and followup ground-based
near-infrared imaging (J, H and K) of the HDF (Dickinson et al
1999). From this catalog we extract the 74 galaxies with spectroscopic
redshifts, $z<1.35$, and with broadband optical and near infrared
magnitudes (a total of 7 passbands). The upper redshift limit is
imposed on our galaxy selection to remove the effect of the IGM on the
observed colors of the galaxies (which comes into effect at
$z>2$).

An other reason to limit most of our study to this redshift range is 
that the number of galaxies with known redshift and reliable photometry is 
small. Using higher redshifts would force us to extend the wavelength 
range of the reconstructed templates on the low wavelength end. This part 
of the templates would be estimated based on the information from a 
small number of high redshift galaxies (usually with larger photometric errors) 
resulting larger errors in the shape of the templates.

For the deconvolution we assume the filter curves and CCD
quantum efficiency curves that are publicly available at the STScI and
National Optical Astronomical Observatories websites. We note in
passing that the resolution of the quantum efficiency curve of the
infrared camera (IRIM) used for the near infrared observations is
poorly known (i.e.\ the resolution is poor) and that a better
representation of this curve may improve on the accuracy of the
results we present below.

Using these data we apply the optimization algorithm described in
Section 2. The redshift range of the data coupled with the filter set
enables us to reconstruct the spectral energy distributions over a
wavelength range of 976--21946 \AA. Within this interval we
reconstruct 3 eigenspectra described by 20 Legendre polynomials
(sampled at a resolution of 450 \AA). The resultant spectral energy
distributions, after approximately 50 iterations, are shown in Figure
6. The solid line shows the first eigenspectrum and the dashed and
dotted lines the second and third eigenspectra respectively. The
relative flux error (the deviation between the colors we would
estimate based on the three eigenspectra and those measured for each
galaxy) is less than 9\%. It is quite remarkable that with only 3
eigenspectra and a poor spectral resolution we could reconstruct the
fluxes with errors that are comparable with the errors in the
observations.

The limit on the resolution of the reconstruction is the number of
galaxies with redshifts and multicolor photometry. For the 74 galaxies
within the HDF sample, observed in 7 passbands, there are a total of
296 degrees of freedom (for 3 eigenspectra - see Equation 7). The 20
Legendre polynomials used in each of the 3 eigenspectra add up to a
total of 60 parameters that we must solve for. The current solution
is, therefore, well constrained. With a larger sample of galaxies (as
we describe in Section 2.1) we could improve on the spectral
resolution. 

Even given the poor resolution of the reconstructed spectra ($\sim
450$ \AA) there are a number of noticeable spectral features present
within the reconstructed data. The first eigencomponent (essentially
the mean of the galaxy distribution) has the spectral shape of a Sbc
or Scd galaxy (cf.\ the Coleman Wu and Weedman galaxy spectral energy
distributions). At the redshifts we probe ($\bar{z} = 0.8$) this is
consistent with the median galaxy type (Lilly et al 1995). In the
first eigenspectrum there is also clear evidence for a break in the
galaxy spectrum at 4000 \AA\ (due to Balmer series absorption). The
second and third eigenspectra appear to be dominated by emission in
the ultraviolet part of the spectrum with a strong upturn at $\lambda
< 4000$ \AA, consistent with a star forming population. In fact the
distribution of the spectral continuum shape of the eigenspectra are
consistent with those we derive from both the BC96 models and from
observations of local galaxies.

For wavelengths greater than 1.4 $\mu$m the reconstructed spectra
suffer from a ringing in the reconstruction. As we noted previously,
because the information used in reconstructing this spectral region is
derived only from the lowest redshift galaxies within our sample then
the long wavelength regions will have lower signal-to-noise.
Consequently these long wavelength regions are more susceptible to the
limited number of galaxies used in the reconstruction (i.e.\ the first
signs of the error show up here). We note, however, that when
calculating magnitudes from these spectral templates the convolution
with the filters partially averages out these fluctuations.

\section{Photometric-Redshifts from Empirical Spectral Energy Distributions}

While a comparison between how well we can reconstruct the observed
colors of galaxies using the empirical spectral templates is a measure
of the ``goodness-of-fit'' the goal of this technique is to improve
the accuracy of the photometric redshift relation. We, therefore,
compare the photometric redshift-relation we derive using these
empirical eigentemplates with those relations given in the literature
(Sawicki et al 1996, Gwyn and Hartwick 1996, Fernandez-Soto et al
1999). Principally we concentrate on the redshift estimators of
Fernandez-Soto et al (1999), for which we have an identical set of
photometric observations.

Using the eigentemplates as spectral energy distributions we use the
standard template fitting method to derive a sample of photometric
redshifts. A comparison between the estimated and spectroscopic
redshifts, for the redshift range $0<z<1.35$, is given in Figure 7. The
solid line represents a one-to-one comparison between the photometric
and spectroscopic redshifts. About this line the rms dispersion within
the relation is $ \sigma_z = 0.077 $ using three eigencomponents. This
compares favorably with the results of Fernández-Soto et al.\ (1999)
who achieve a dispersion of $ \sigma_z = 0.095 $ using the four
Coleman, Wu and Weedman (1980) (CWW) spectral energy distributions and also
with the polynomial fitting techniques of Connolly et al.\ (1995) who
determine an $ \sigma_z = 0.14 $ and $ \sigma_z =0.062 $ with first and 
second order polynomial fits respectively.

Although the number of galaxies with $z>1.5$ is small to get
reliable templates for the high redshift galaxies we made a test
to demonstrate that the method does not brake over this redshift 
limit (Figure 8.). The higher redshift range increase the number of 
observations to $102$ from the original $74$ but the wavelength range needed 
for the templates also extends. 
Since  higher redshifts strech the restframe spectra more,
to get the same results the resolution of templates should increase also.
The number of objects with $z>1.5$ is small, so these requirements can be 
satisfied only if we reduce the number of eigenspectra from $3$ to $2$.
The comparison of the spectroscopic and photometric redshifts can be seen on 
Figure 8. There is one extreme outlier and there seems to be some systematic 
underestimation of redshift for high $z$ objects. Despite of these errors the
redshift estimation is better than the one with the CWW templates. The rms 
error is  $ \sigma_z = 0.34 $ with our estimated and $ \sigma_z = 0.40 $ with 
the CWW templates. The few outliers are responsible for most of the error.
If we remove the one (three) extreme outlier(s) with $\Delta z > 1$ we got 
$ \sigma_z = 0.17 $ ($ \sigma_z = 0.22$) rms dispersion with the estimated 
(CWW) templates, respectively. 

\section{Discussion and Applications}

The technique we described above incorporates the strengths of the
empirical and template based photometric redshift techniques.  It does
this by utilizing a training set of galaxies (with colors and
redshifts) to derive a set of spectral energy distributions (as
opposed to defining a general but somewhat arbitrary set of polynomial
coefficients). The result of this optimization procedure is to produce
galaxy spectra that have been optimized to match the color
distribution of galaxies within a given sample. It is important to
note that the spectra are not optimized to produce the spectroscopic
redshifts of the training set and, therefore, a comparison between
the observed (spectroscopic) and predicted (photometric) redshifts is
a statistically fair comparison. As we have defined a physical basis
for the photometric-redshift relation (as opposed to the general
polynomial relation) we can apply the spectral energy distributions
over a wide range in redshift and are not restricted to just the
redshift range over which we derive the relation (which was a
fundamental limitation on our earlier approach).

As we have shown in Figure 7 applying this technique we reduce the
scatter about the photometric-redshift relation to $\sigma_z = 0.077
$. This compares favorably with the results obtained by Fernandez-Soto
et al (1999) who obtain a dispersion of $\sigma_z = 0.095 $ for
galaxies with $z<1.35$.  The reason for the decrease in the dispersion
about the photometric-redshift relation is the improvement in the
spectrophotometric templates used for estimating the galaxy
redshifts. The standard Coleman, Wu and Weedman (1980) templates,
while producing a remarkably good fit, are based on the spectra of
approximately 12 {\it local} galaxies. The optimization technique we
employ utilizes the colors of 74 galaxies (distributed over a range of
redshifts) and is, therefore, more likely to sample the full
distribution of galaxy types.

The limitation of our current application is simply the number of
galaxies with accurate multicolor photometry in the Hubble Deep
Field. The optimization technique is, however, general enough that we
can incorporate any multicolor photometric and spectroscopic survey
into the analysis. With the new generation of multicolor redshift
surveys nearing completion we expect this approach to substantially
improve on standard template-based photometric-redshift relations over
the next few years. It should also be noted that as the result of this
analysis is also a set of spectral energy distributions (or more
exactly a statistical representation of the spectra) the eigenspectra
can be applied to new multicolor surveys without requiring that they
be transformed to the same photometric system.

As the spectral energy distributions we reconstruct are derived
directly from observations they include the effects of dust and galaxy
evolution. It is reasonable to expect that these spectra will be a
better representation of galaxies over a wide range in redshift than
standard local galaxy templates (as noted above). The derived
eigentemplates can therefore be used to construct a set of
K-corrections optimized for galaxies over a wide range in
redshift. Ultimately, with large photometric samples, we can take this
analysis one step further. Comparing the predictions of spectral
synthesis models with those derived from the multicolor photometry
will show how star formation and reddening by dust couple to produce
the observed colors of galaxies. It should, therefore, be possible to
identify where the models of star formation and the observed spectral
properties of galaxies deviate and thereby improve on the spectral
synthesis codes.

Finally we note that the analysis we have currently undertaken
requires the use of multicolor photometry on galaxies of known
spectroscopic redshift. On can extend this
analysis to the case of data with only multicolor observations (i.e.\
no spectroscopic redshifts). We could optimize the estimated fluxes
not only for the expansion coefficients and for the shape of the
eigenspectra as we described above but also for the redshifts.  This
would naturally increase the sample of galaxies available for the
optimization procedure (enabling much finer resolution for the
resultant eigenspectra). It would also become a correspondingly larger
computational problem. Another possibility would be to use
photometric rather than spectroscopic redshifts.

\section{Conclusions}

We have presented a new technique that can reconstruct the continuum
spectra of galaxies directly from a set of multicolor photometric
observations and spectroscopic redshifts. Using simulated multicolor
data we show that we can recover the underlying spectral energy
distribution even in the presence of substantial amounts of
noise. Applying this approach to existing optical and near-infrared
photometric data from the Hubble Deep Field we derive a set of
spectral energy distributions that describe the observed galaxy
colors. The main spectral features present in the spectral energy
distributions of galaxies could be clearly seen within the
reconstructed low resolution eigenspectra. The utility of this
approach is demonstrated by using the empirical spectral energy
distributions in a template-based photometric-redshift relation. The
photometric redshift estimation based on the resultant template
spectra gives redshift errors significantly better than standard
template techniques. The current limitation on the accuracy of our
technique is simply the number of high signal-to-noise multicolor
photometric data currently available.  Given the new photometric and
spectroscopic surveys underway or nearing completion we anticipate a
significant improvement in the resolution and accuracy of the derived
spectral energy distributions.

\acknowledgments

We would like to thank Daniel Eisenstein, Jim Annis and David Hogg for
useful discussions about our reconstruction technique. IC acknowledges
partial support from the MTA-NSF grant no. 124 and the Hungarian
National Scientific Research Foundation (OTKA) grant no.\ T030836, AJC
acknowledges support from an LTSA grant (NAG57934). AS acknowledges
support from NSF (AST9802980) and a NASA LTSA (NAG53503).

\clearpage
\begin{figure}
\centerline{\hbox{\psfig{figure=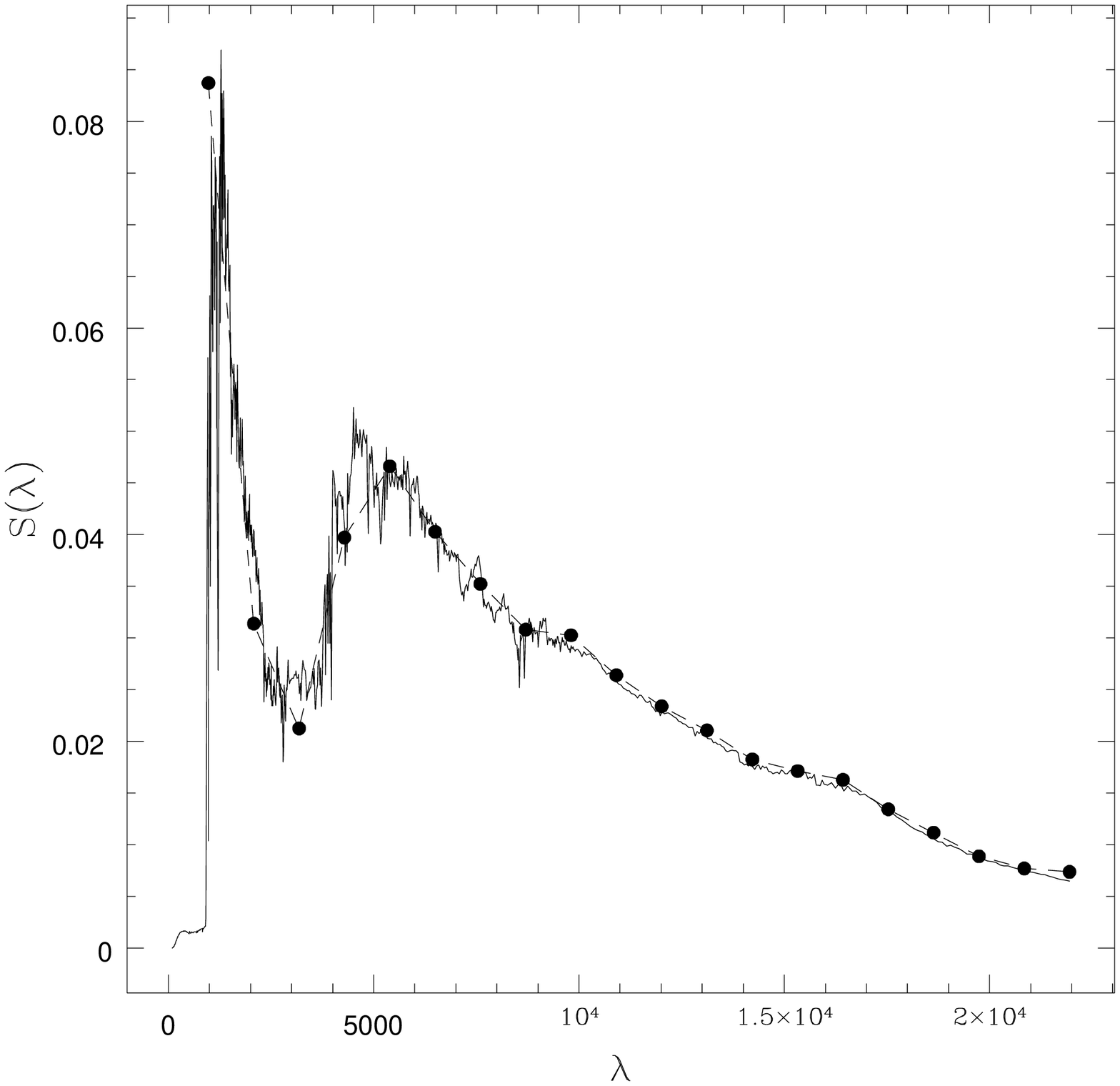,height=3.5in}
\psfig{figure=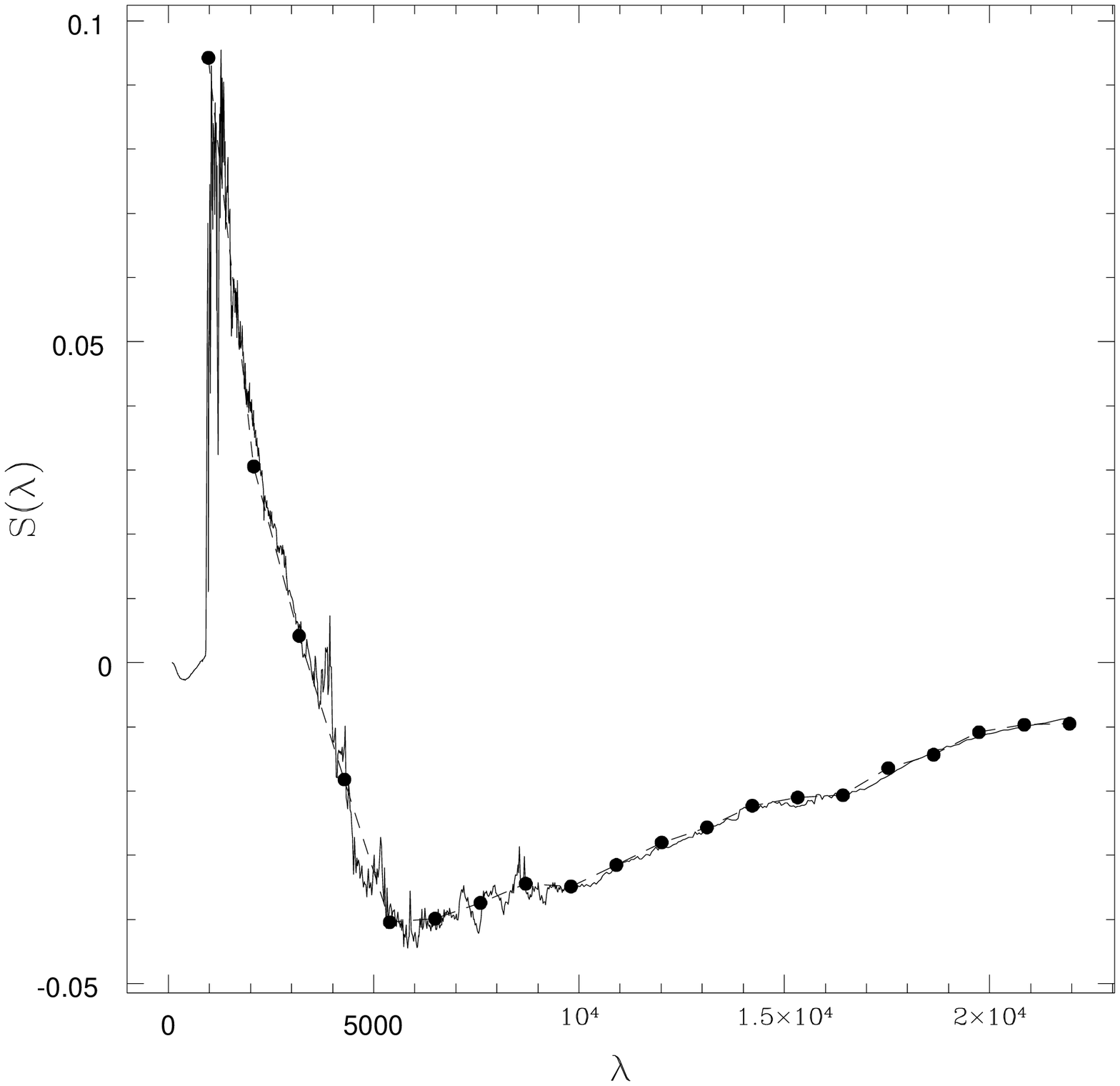,height=3.5in}}}
\caption{The left hand panel shows the first eigenspectrum (solid
line) used to generate a simulated photometric catalog. The filled
circles represent the reconstructed eigenspectrum derived from the
multicolor photometry. The reconstruction was undertaken using 20
Legendre polynomials as the basis functions sampled at 20 points
(i.e.\ with a resolution of approximately 1000 \AA). The expansion
coefficients, $b_{jl}$ (see Equation 3), for the Legendre polynomials
were were initialized with random values. The reconstructed spectrum
shown above was achieved after 50 iterations. The right hand panel
shows the second eigenspectrum and the corresponding second
reconstructed eigenspectrum. The rms deviation between the original
and reconstructed eigenspectra is 6.98\% and 6.99\% for the first and
second eigenspectra respectively}
\end{figure}

\begin{figure}
\centerline{\hbox{\psfig{figure=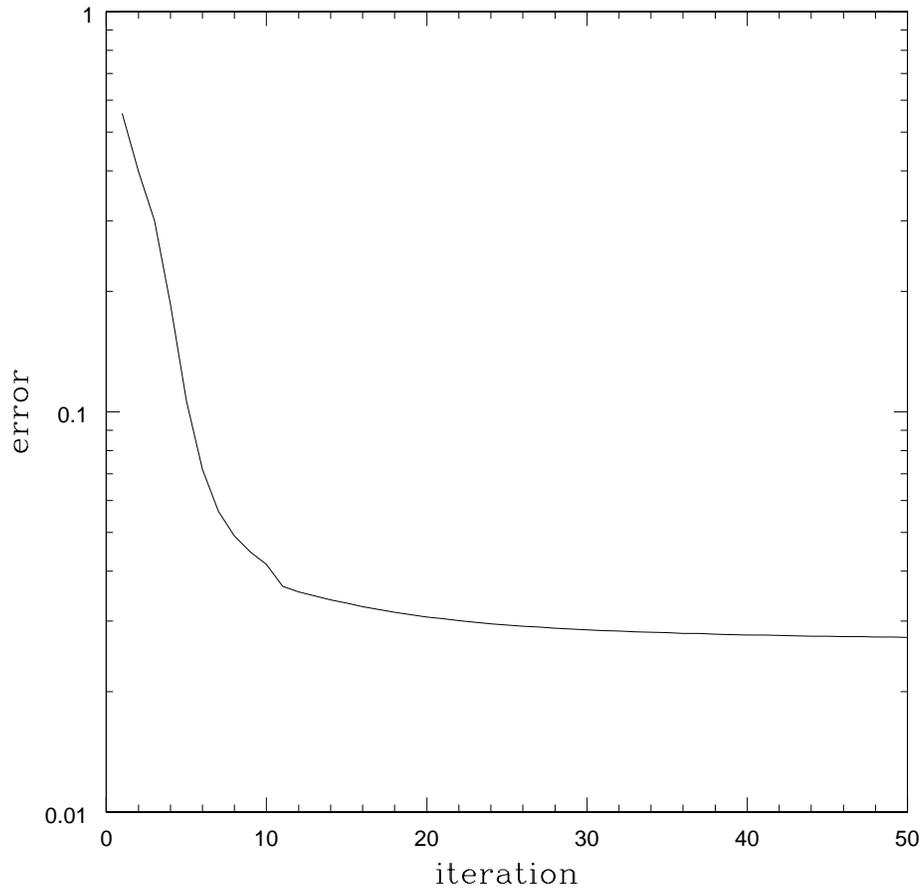,height=5in}}}
\caption{The convergence of the relative error as a function of the
number of iterations for the reconstruction of the galaxy
eigenspectra.  The relative error decreases rapidly and by the 50th
iteration varies by $<0.1$\% from one iteration to the next.}
\end{figure}
\begin{figure}

\centerline{\hbox{\psfig{figure=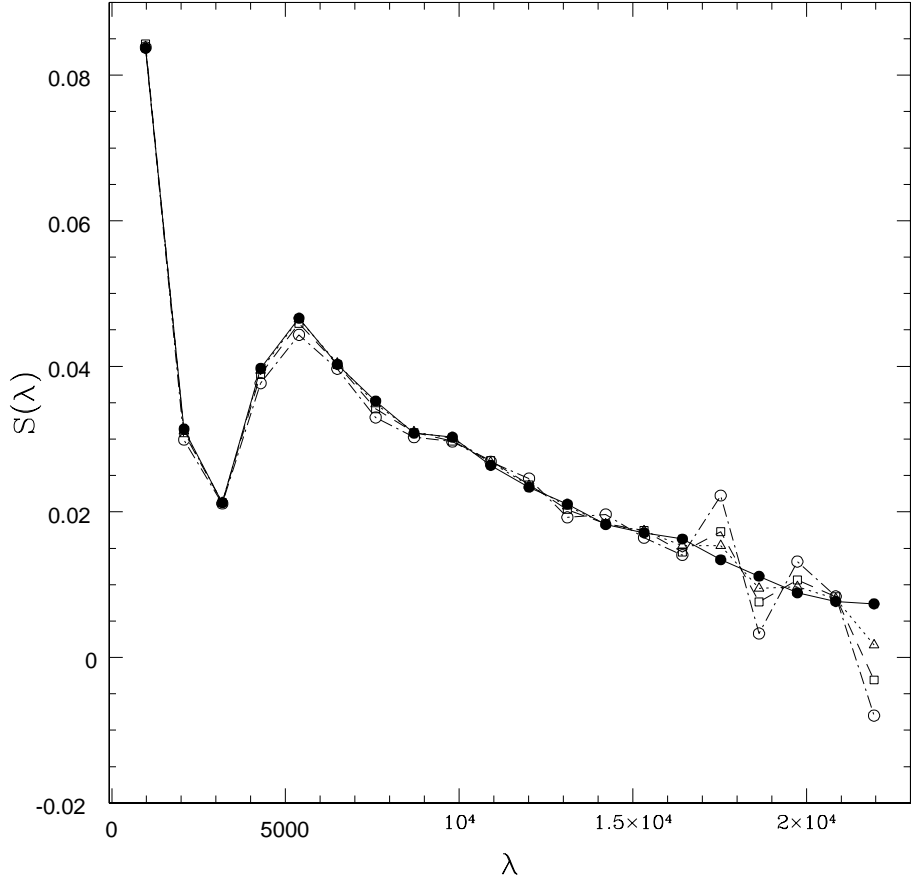,height=5in}}}
\caption{The effect of photometric uncertainty on the reconstruction
of the first eigenspectrum. The filled circles show the reconstruction
if there were no photometric errors within the data. The triangles,
squares and open circles show the effect of increasing photometric
errors within the data to 5\%, 10\% and 20\% respectively.  With a
large number of galaxies the underlying eigenspectra that describe the
galaxy distribution can be reconstructed to a high degree of accuracy
even with relatively low signal-to-noise data. The multiplex advantage
of having many galaxies sampling the same spectral intervals means
that we have a resultant increase in the overall signal-to-noise of
the output spectra (i.e.\ we are essentially coadding the data to beat
down the noise). Since the long tails of the spectra are reconstructed
using information only from highest and lowest redshift galaxies, the
first signs of the error show up there. When calculating magnitudes
the convolution with the filters partly averages out these
fluctuations.}
\end{figure}

\begin{figure}
\centerline{\psfig{figure=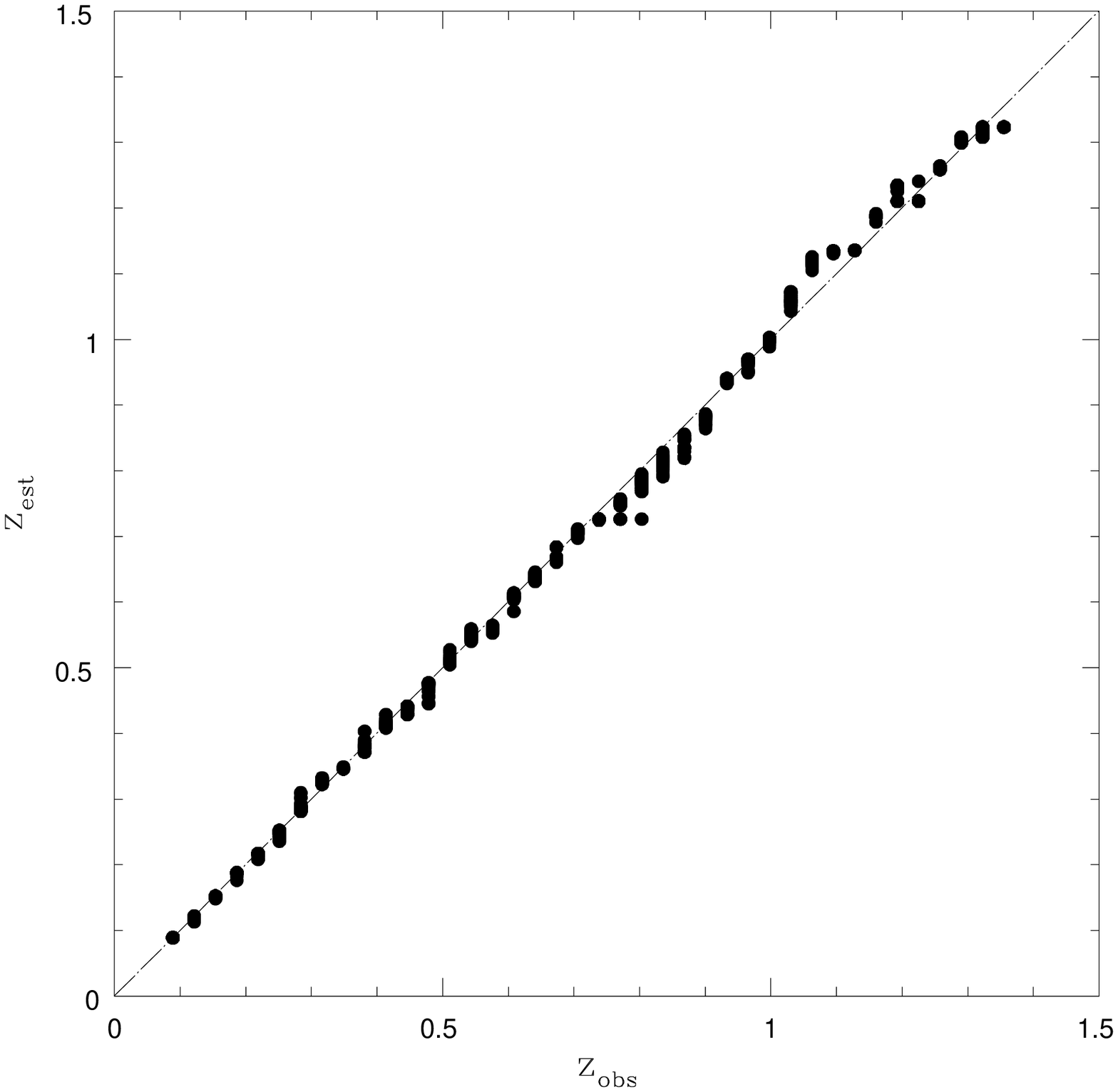,height=3in}
\psfig{figure=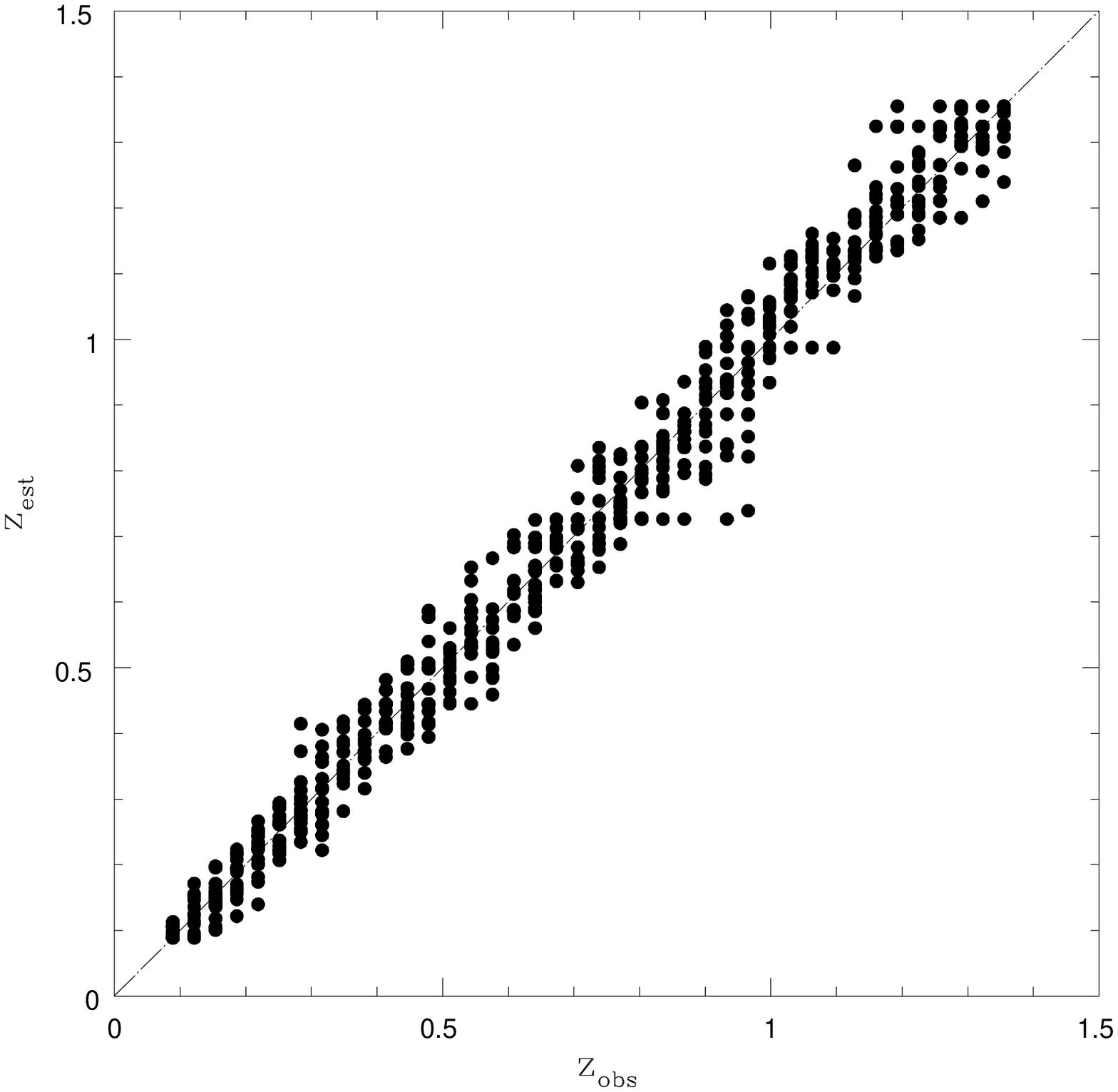,height=3in}} 

\centerline{\psfig{figure=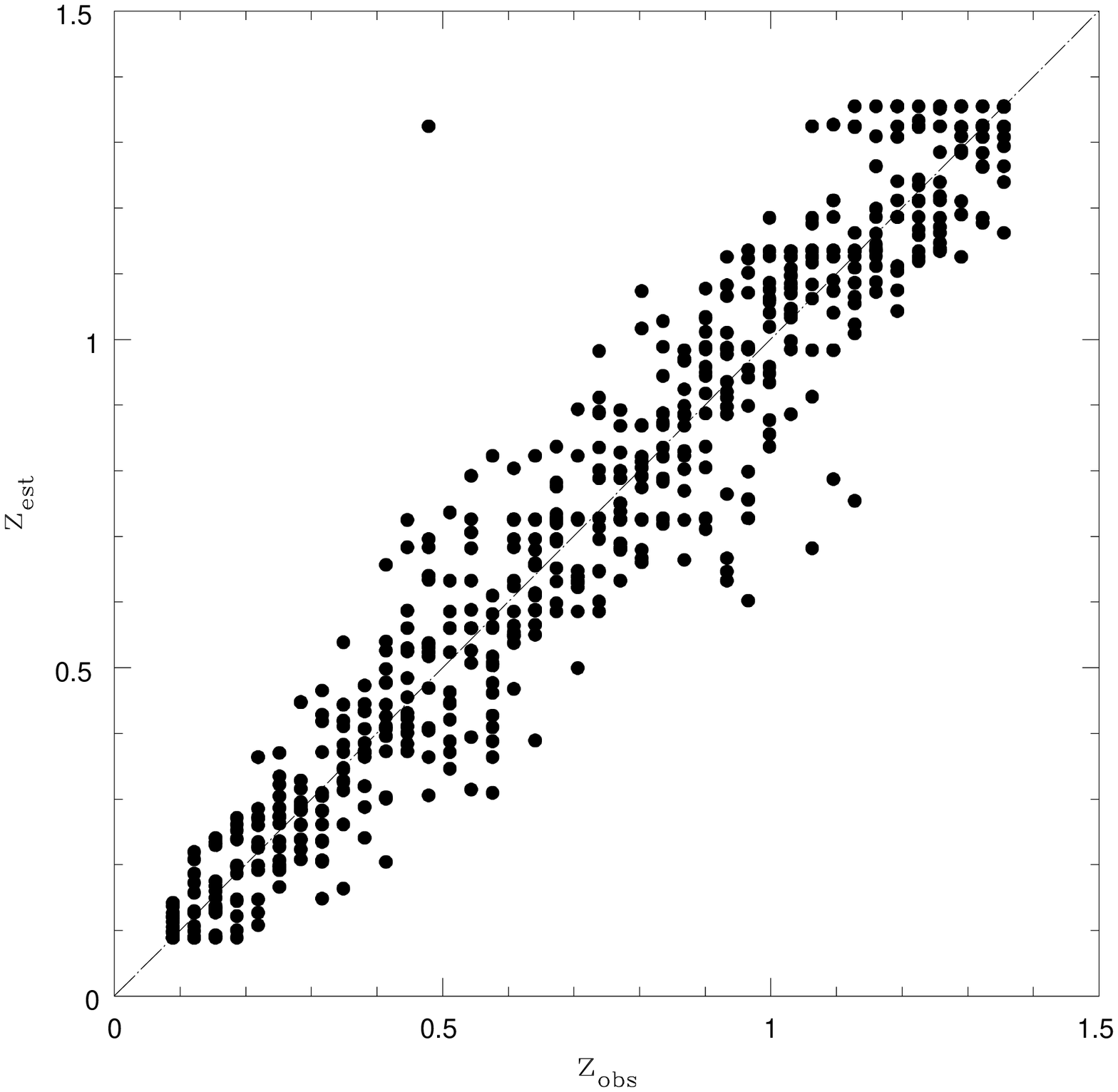,height=3in}
\psfig{figure=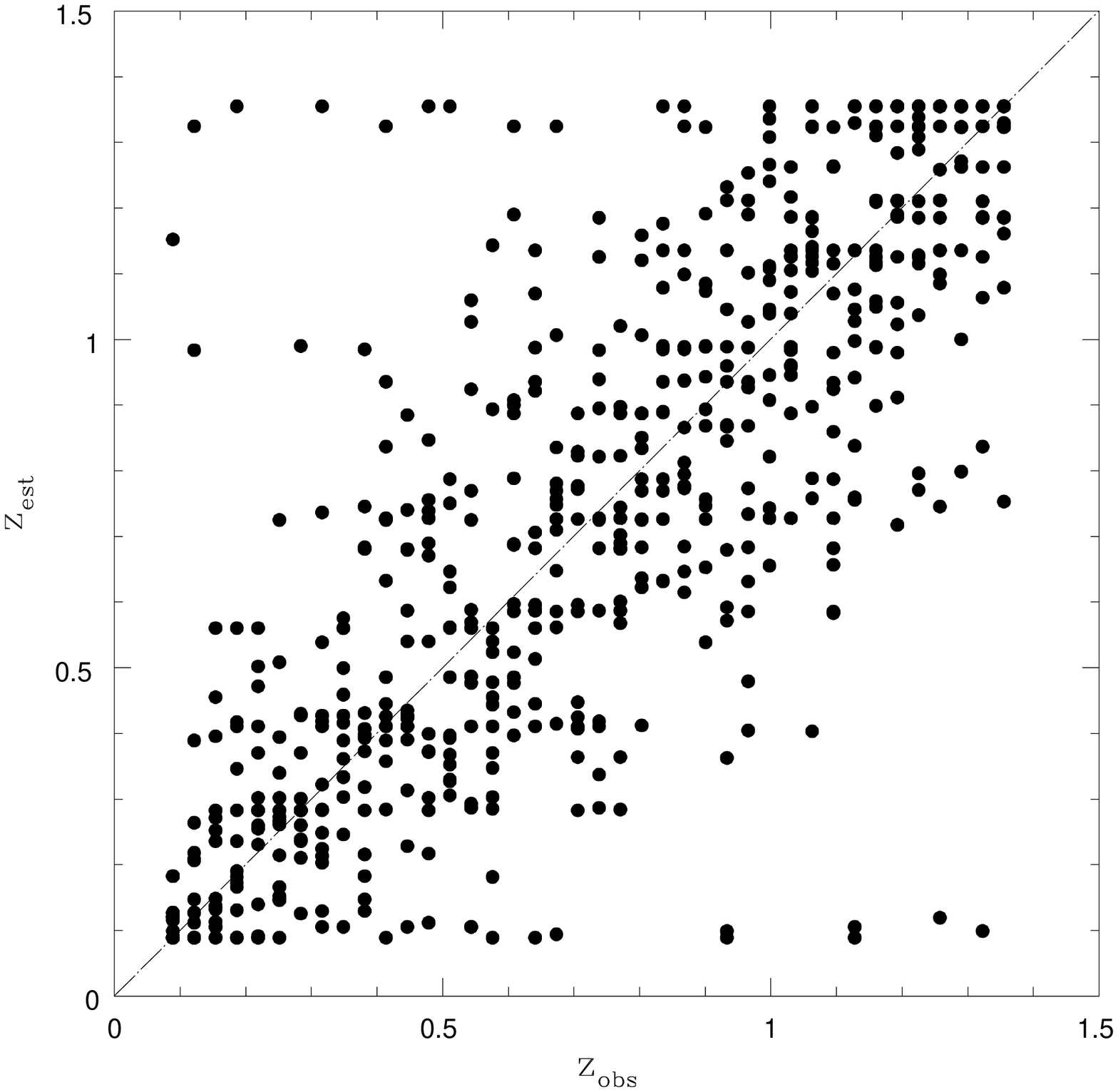,height=3in}}
\caption{ The photometric-redshift relation derived from the simulated
data set using a range of signal-to-noise ratios. The panels, from
left to right and top to bottom, show the effect of 0\%, 5\%, 10\% and
20\% errors in the photometric data. For the analysis it was assumed
that each of the photometric passbands had an identical
signal-to-noise.}
\end{figure}

\begin{figure}
\centerline{\psfig{figure=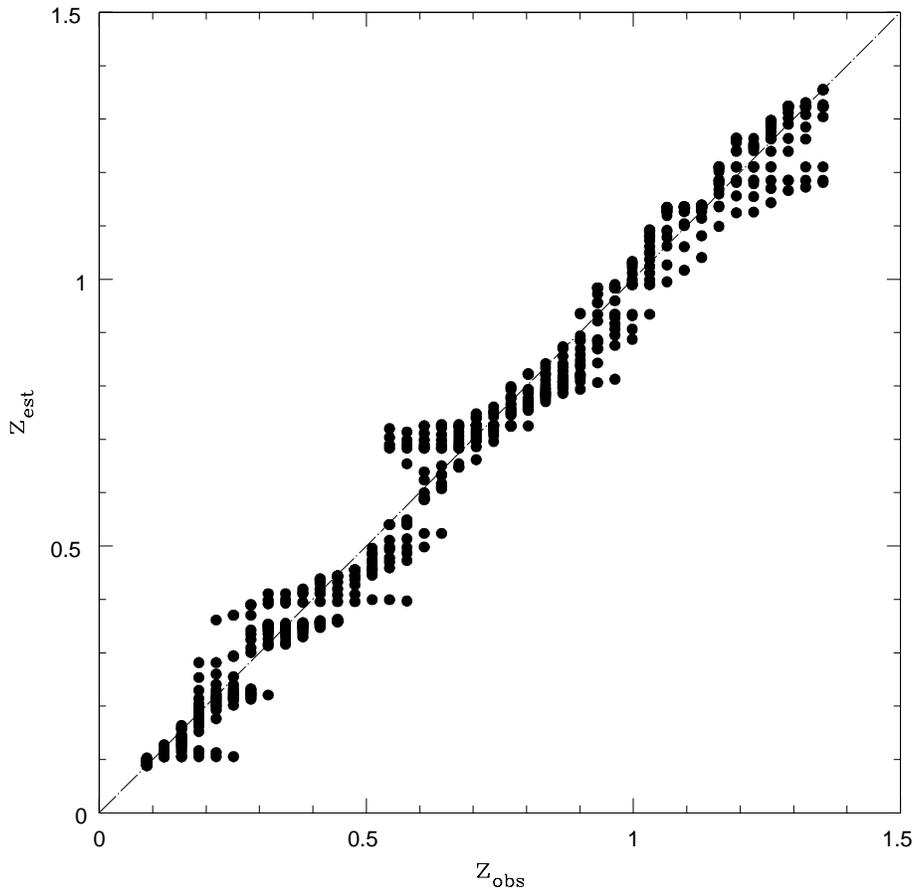,height=5in}}
\caption{The photometric redshift relation for the simulated data set
where the eigenspectra have been defined using a set of photometric
data with 20\% errors in the flux and then applied to a data set with
0\% error in the flux. The dispersion about this relation is $\sigma_z
= 0.05$ which is comparable to the $\sigma_z = 0.02$ found for the
ideal dataset. This validates the analysis shown in Figure 3 where,
even with large photometric errors, we can reconstruct the underlying
eigensystem with a very high degree of accuracy. The uncertainty in
the resultant photometric redshift relation is, therefore, dominated
by the signal-to-noise of the data we wish to apply it to. }
\end{figure}

\begin{figure}
\centerline{\hbox{\psfig{figure=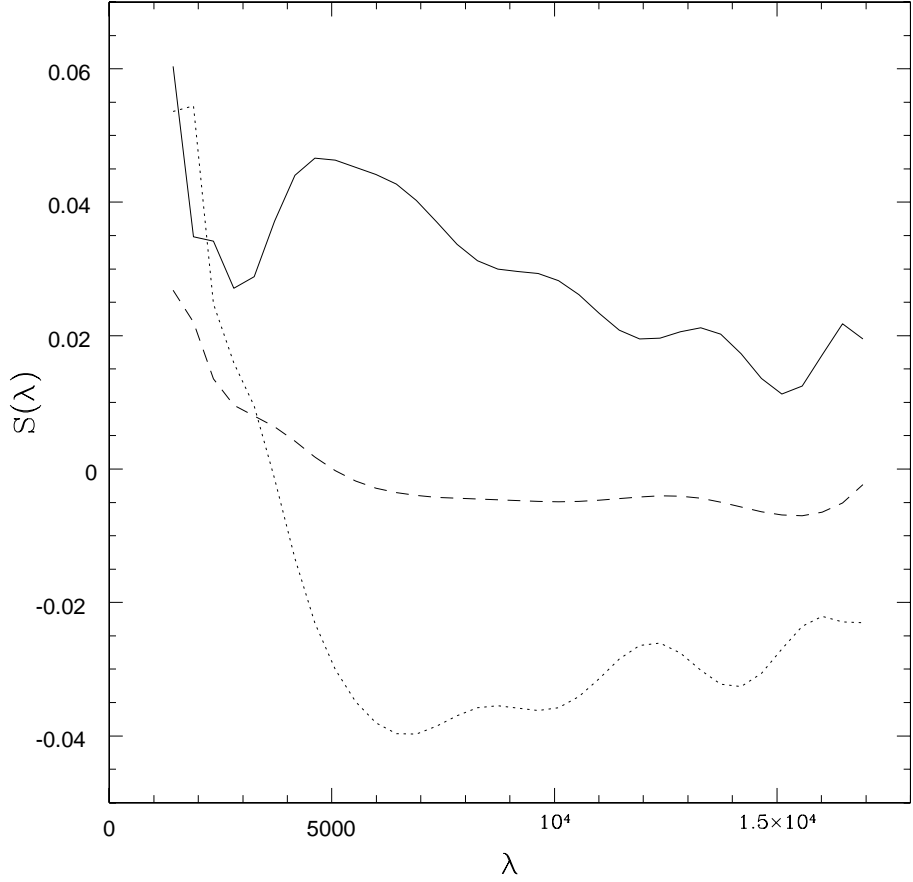,height=5in}}}
\caption{ The HDF catalog with 74 galaxies $z<1.35$ were used to
reconstruct spectral templates. The eigentemplates were iterated from
random parameters using an orthogonal base of 20 Legendre polynomials
(sampled at 450 \AA\ resolution). The solid line shows the first
eigenspectrum derived from the data and the dashed and dotted lines
the second and third eigenspectra respectively.  Even given the low
resolution nature of the output spectra the break at 4000 \AA\ and the
rise in the ultraviolet continuum due to the presence of star
formation are clearly visible. }
\end{figure}

\begin{figure}
\centerline{\hbox{\psfig{figure=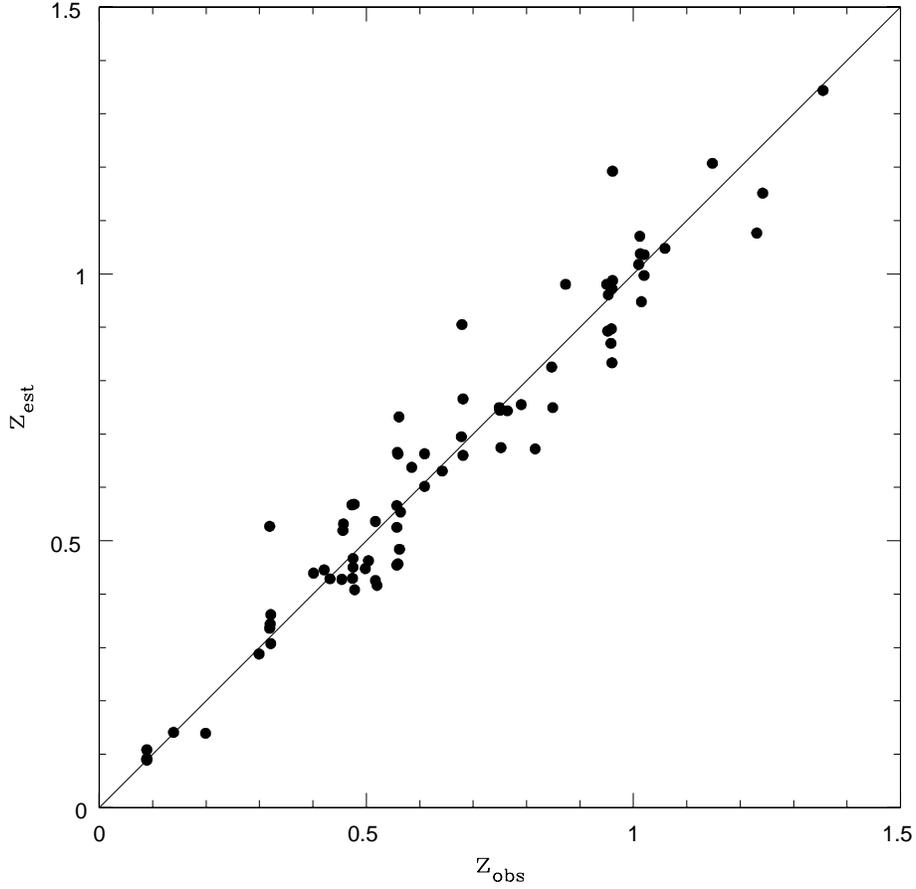,height=5in}}}
\caption{ The photometric redshift estimation using the three
reconstructed eigenspectra from the HDF photometric catalog (see
Figure 4). This analysis gives a smaller dispersion in the photometric
redshifts relation, $\sigma_z = 0.077$, than that derived by
Fernandez-Soto et. al. (1999), $\sigma_z = 0.095$, using the four
Coleman, Wu and Weedman (1980) (CWW) model/empirical spectral energy
distributions. Polynomial photometric redshift estimation technique
(Connolly et al 1995) gave $\sigma_z = 0.14$ and $\sigma_z = 0.062$ with first 
and second order polynomials respectively.
}
\end{figure}

\begin{figure}
\centerline{\hbox{\psfig{figure=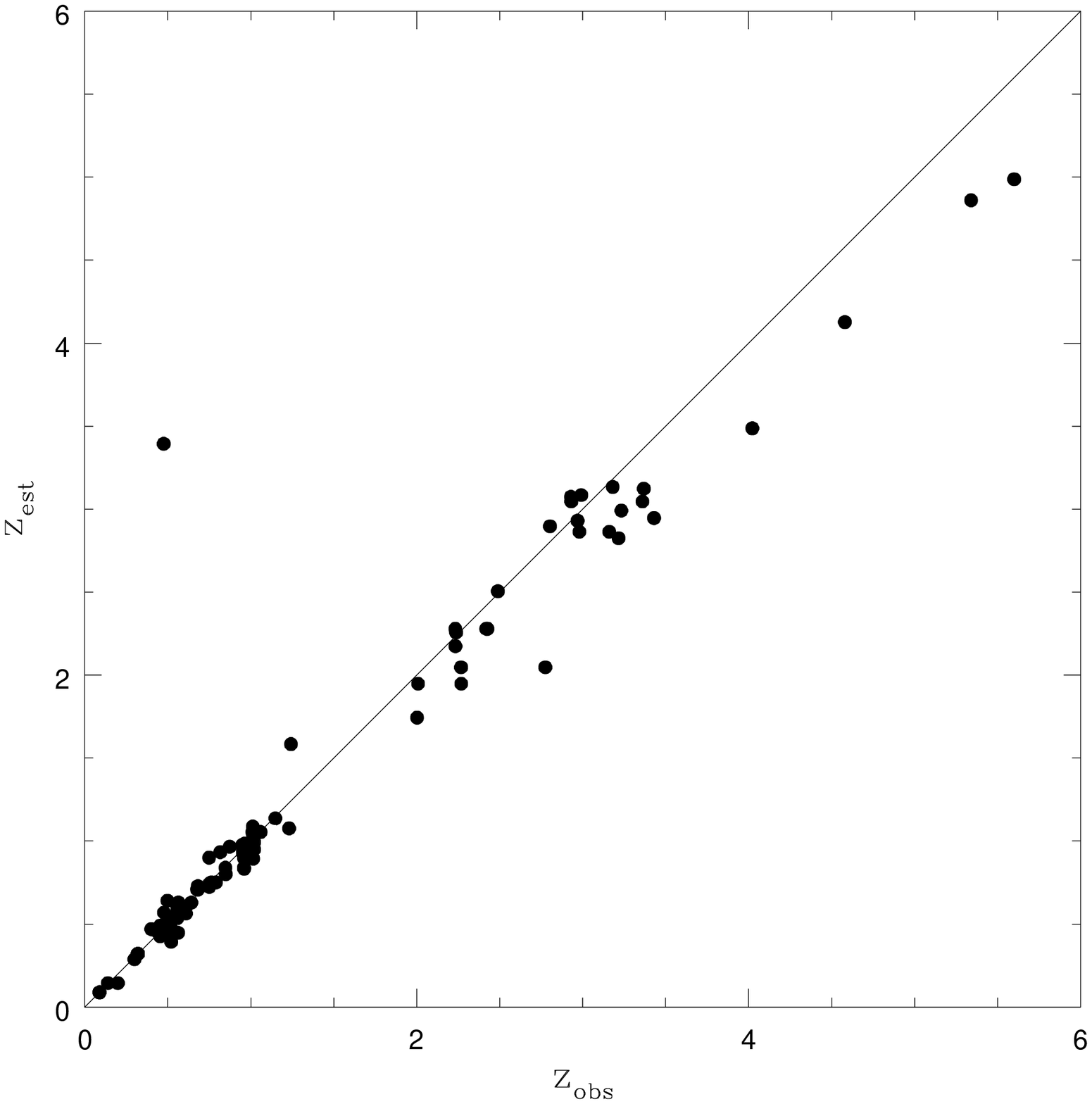,height=5in}}}
\caption{ The photometric redshift estimation using two 
reconstructed eigenspectra from the HDF photometric catalog including
high redshift objects also. 
Despite of the small number of high redshift objects in the training set the 
estimated templates give better photometric redshifts
($\sigma_z = 0.34$), than that derived by
Fernandez-Soto et. al. (1999), ($\sigma_z = 0.40$), using the CWW 
spectral energy distributions. 
Removing the extreme outliers ($\Delta z > 1.0$) 
the dispersion decreases to $\sigma_z = 0.17$ and $\sigma_z = 0.22$ 
for the estimated and CWW templates respectively.
}
\end{figure}

\appendix


\begin{references}

\reference{} Benitez, N., 1999, astro-ph/9811189, ApJ in press.
\reference{} Bruzual, A.G. \& Charlot, S., 1993, ApJ, 405, 538
\reference{} Budav\'ari, T., Szalay, A.S., Connolly, A.J., Csabai, I. \&
Dickinson, M.E., 1999, in preparation 
\reference {} Coleman, G.D., Wu., C.-C. \& Weedman, D.W., 1980, ApJS, 43, 393
\reference{} Connolly, A.J., Csabai, I., Szalay, A.S., Koo, D.C.,
Kron, R.G. \& Munn, J.A., 1995a, AJ 110, 2655
\reference{} Connolly A.J., Szalay A.S., Bershady M.A., Kinney A.L. \&
Calzetti D., 1995b, AJ, 110, 1071
\reference{} Dickinson, M.E. et al, 1999, in preparation
\reference{} Fernandez-Soto, A., Lanzetta, K.M. \& Yahil, A., 1999, in press
\reference{} Gwyn, S.D.J. \& Hartwick, F.D.A., 1996, ApJ, 468, L77
\reference{} Karhunen, H., 1947, Ann. Acad. Science Fenn, Ser. A.I. 37
\reference{} Lilly, S.J., Tresse, L., Hammer, F., Crampton, D. \& Le Fevre, O.,
1995, ApJ, 455, 108
\reference{}  Lo\`{e}ve, M., 1948, Processus Stochastiques et Mouvement Brownien, Hermann, Paris, France
\reference{} Madau, P., Ferguson, H.C., Dickinson, M.E., Giavalisco,
M., Steidel, C.C. \& Fruchter,A., 1996, MNRAS, 283, 1388
\reference{} Sawicki, M.J., Lin, H. \& Yee, H.K.C., 1997, AJ, 113, 1
\reference{} Williams, R.E. et al., 1996, AJ, 112, 1335
\end{references}
\end{document}